\documentclass{iaus}
\usepackage{graphics}

  \checkfont{eurm10}
  \iffontfound
    \IfFileExists{upmath.sty}
      {\typeout{^^JFound AMS Euler Roman fonts on the system,
                   using the 'upmath' package.^^J}%
       \usepackage{upmath}}
      {\typeout{^^JFound AMS Euler Roman fonts on the system, but you
                   dont seem to have the}%
       \typeout{'upmath' package installed. iaus.cls can take advantage
                 of these fonts,^^Jif you use 'upmath' package.^^J}%
      }
  \else
  \fi

  \checkfont{msam10}
  \iffontfound
    \IfFileExists{amssymb.sty}
      {\typeout{^^JFound AMS Symbol fonts on the system, using the
                'amssymb' package.^^J}%
       \usepackage{amssymb}%
         \let\leq=\leqslant
         
      }{}
  \fi

  \IfFileExists{amsbsy.sty}
    {\typeout{^^JFound the 'amsbsy' package on the system, using it.^^J}%
     \usepackage{amsbsy}}
    {\providecommand\boldsymbol[1]{\mbox{\boldmath $##1$}}}

\newsavebox{\astrutbox}
\sbox{\astrutbox}{\rule[-5pt]{0pt}{20pt}}

\title[Outskirts of Galaxy Clusters: intense life in the suburbs]
      {Evolutionary Status of Early--Type Galaxies in Distant Poor Clusters}

\author[A. Fritz \& B.~L.\ Ziegler]{Alexander Fritz$^1$ \and Bodo L.\ Ziegler$^1$}

\affiliation{$^1$Universit\"ats--Sternwarte G\"ottingen, Geismarlandstra{\ss}e 11,
37083 G\"ottingen, Germany email: afritz@uni-sw.gwdg.de\\[\affilskip]}

\pubyear{2004}
\volume{195}
\pagerange{1--8}
\date{?? and in revised form ??}
\jname{Outskirts of Galaxy Clusters: intense life in the suburbs}
\editors{A. Diaferio, ed.}
\begin{document}

\maketitle

\begin{abstract}
We introduce our project that investigates the kinematic properties of
early--type galaxies in 6 distant poor clusters at $z\approx0.25$. This
study represents a continuation of our efforts to understand galaxy
evolution in low-density environments. Higher--resolution MOSCA spectra have
been obtained at the Calar Alto 3.5-m telescope with which we can measure
absorption line strengths and velocity dispersions. In conjunction with our
\textit{HST/F702W} images of all the clusters, we are able to construct the
Fundamental Plane of ellipticals and S0 galaxies in poor clusters at a
look-back time of $\approx3$\,Gyrs. For galaxies outside the \textit{HST}
field, we concentrate our analysis on the Mg--$\sigma$ and Faber--Jackson
relations. With the line strength diagrams age/metallicity distributions can be
derived in densities between the field and rich cluster environments. Comparing
with our rich clusters at the same cosmic epochs, the dependence of galaxy
formation models on the local environment can be tested more quantitatively.
\end{abstract}

\firstsection 
\section{Introduction}

Recent analyses of the kinematics and stellar populations of faint distant
galaxies up to redshifts of $z\approx1.0$ have put strong observational
constraints on galaxy formation models. The weak evolution seen in the tight
correlation of the distance independent parameters Mg--absorption and velocity
dispersion, $\sigma$, at $z\approx0.4$ is evidence for a high redshift
($2<z_f<4$) of formation for the bulk of the stars in cluster ellipticals
(e.g., \cite{ZB97}). Further support gives the small increase of the
$B$-band luminosity derived from the Faber--Jackson relation
($L$ \textit{vs.} $\sigma$). The combination of structural (size and surface
brightness) and kinematic ($\sigma$) parameters in the Fundamental Plane
(\cite{DD87}) let us investigate not only the luminosity but also
the mass evolution of early--type galaxies. Most previous studies of the
Fundamental Plane (FP) at $0.2<z\leq1.3$ revealed that the mass-to-light ratio
of luminous galaxies evolves only mild in accordance with passive evolution
models (\cite{Z2001}; \cite{vDS03}; see also contribution by
Fritz \& Ziegler in this volume). However, these observational results apply
only to the special environment of dense rich clusters.

Semi-analytic CDM models predict a hierarchical assembly for
galaxies through merging of subunits and replenishment of gas from their halos
(\cite{Col:00}). Only in the richest clusters galaxies experienced their
last major merger at $z\sim1-2$. Thus, the population of early-type galaxies
should be more diverse in lower density regions.

For this reason, we analyse the evolutionary status of early--type (E+S0)
galaxies in poor clusters at a look-back time of $\approx3$\,Gyrs with the same
quantitative methods as was done in rich clusters. The Mg--$\sigma$ relation
will reveal whether there is more spread in the age/metallicity distribution
than in rich clusters at similar epochs, e.g., in Abell\,2218
(\cite{Z2001}) or Abell\,2390 (\cite{Fri:04}). Results on the FP will give
insights whether the mass/structure evolution is more consistent with the
monolithic or the hierarchical formation scenario. We use
$\Omega_{m}=0.3$, $\Omega_{\Lambda}=0.7$ and $h^{-1}_{70}$.

\section{The Low-$L_{X}$ Project}

This investigation is an intensive observational campaign to study the
evolutionary status of galaxies in 10 poor clusters at intermediate redshifts
of $0.2<z<0.3$ selected to have very low X-ray luminosities
($L_{X}<5\times10^{43}$~erg/s), 100 times lower than the CNOC clusters, and
poor optical richness class. It comprises ground-based optical and NIR
photometry, medium-resolution multi-object spectroscopy,
and \textit{HST/F702W} imaging.

The clusters are dominated by passive galaxies with no or little on-going
star formation (SF). The fraction of emission-line galaxies (ELG) is as low as
in the rich X--ray bright CNOC clusters at similar epochs
($f_{\rm ELG}\approx0.2$) and at variance with the field value
($f_{\rm ELG}\approx0.6$). This indicates that processes
heavily depending on density like ram--pressure stripping can not be the
dominant mechanisms responsible for the supression of SF (\cite{lowLx}).
Analysis of the \textit{HST} images for all ten clusters revealed that the
overall properties of these cluster galaxies are similar to those in more
massive clusters with a low fraction of disk galaxies (\cite{lowLxhst}).

\section{Sample Selection for Spectroscopy}

The spectroscopic observations have been gained during eleven nights
(04/2001 and 02/2002) with the MOSCA spectrograph at the Calar Alto Observatory
(Spain). A summary of the cluster sample with the mean redshifts, number of
objects on a mask and total exposure times for spectroscopy is given in
Tab.\,\ref{tab:samp}.
Early-type cluster members have been selected based on our low-resolution
spectra and from their \textit{HST} morphology. For the FP study, we obtained
spectra of 5$-$7 galaxies lying within the \textit{HST} fields in a single MOS
mask. Five masks (one per cluster; Cl\,1701 and Cl\,1702 only observed
with one mask) have been observed to construct a combined
FP sample of $\sim$30 galaxies, comparable to that of a single rich cluster
(\cite{Fri:04}). Outside the \textit{HST} field additional spectra of
$\sim$10 E+S0 in each cluster serve for the investigation of the
Mg--$\sigma$ and Faber--Jackson relation.
For these scaling relations the number of observed galaxies is sufficient to
explore possible variations between the different poor clusters as a function
of their $\sigma$ or $L_{X}$. Our total sample comprises $\sim$80 galaxies.

\begin{table}
\begin{center}
\vspace*{-0.4cm}
\caption{Sample of six poor Clusters.}
  \begin{tabular*}{0.83\textwidth}{lcccccc} 
  \hline
  Cluster &  R.A.\  & Dec.\  & $<z>$ &
  $L_{\rm X}$ (0.1--2.4 keV) & $N_{\rm obj/mask}$ & $T_{{\rm exp}}$\\
  & \multispan2{\hfil (J2000)\hfil } & &$10^{43} h^{-2}_{50}$ ergs s$^{-1}$
  & & (ksec)\\
\hline
Cl\,0818+56 & 08 19 04 & +56 54 49 &  0.2670  & 1.50 & 14 & 29.40 \\
Cl\,0841+70 & 08 41 44 & +70 46 53 &  0.2397  & 1.22 & 13 & 24.54 \\
Cl\,0849+37 & 08 49 11 & +37 31 09 &  0.2343  & 1.93 & 17 & 18.00 \\
Cl\,1633+57 & 16 33 42 & +57 14 12 &  0.2402  & 0.49 & 14 & 30.97 \\
Cl\,1701+64 & 17 01 47 & +64 20 57 &  0.2458  & 0.40 & 18 & 27.90 \\
Cl\,1702+64 & 17 02 14 & +64 19 53 &  0.2233  & 0.74 & 18 & 27.90 \\
\noalign{\smallskip}
\end{tabular*}
\label{tab:samp}
\end{center}
\end{table}

\begin{figure}
\begin{center}
\vspace{0.5cm}
\resizebox{7.0cm}{!}{\includegraphics{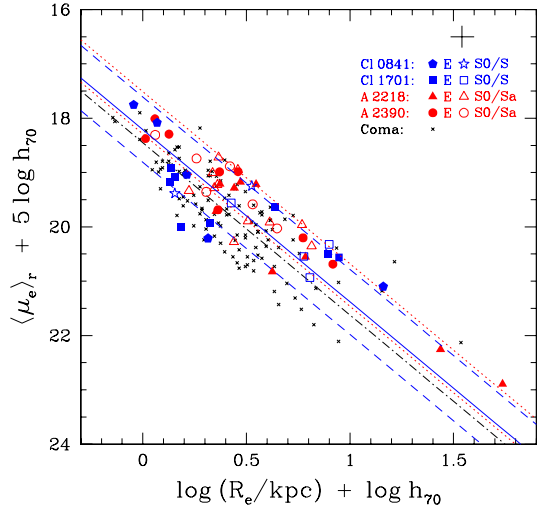}}
\caption{KR in Gunn $r$ for the poor clusters Cl\,0841 ($z{=}0.24$) and
  Cl\,1701 ($z{=}0.25$), compared to the rich clusters A\,2218 ($z{=}0.18$) and
  A\,2390 ($z{=}0.23$) and to the local Coma sample of JFK95.
  The dot-dashed line is a least square fit to Coma. Dotted lines show the
  1~$\sigma$ errors of the fit for the rich clusters, the solid line indicates
  the fit for the poor clusters with 1~$\sigma$ (dashed lines), assuming the
  local slope. A typical error bar is in the upper right corner.}
\label{fig:KR}
\end{center}
\end{figure}

\section{First Results}

A first test on the formation and evolution of elliptical and S0 galaxies
is made with the Kormendy relation (KR), a scaling relation between the
effective radius $R_e$ and effective surface brightness $\langle\mu\rangle_e$.
As this relation is not based on galaxy kinematics, it can be studied to fainter
magnitudes.

The KR for two low $L_{X}$ clusters, Cl\,0841 and Cl\,1701, is shown in
Fig.\,\ref{fig:KR}. We compare these poor clusters with two rich clusters,
A\,2218 and A\,2390 at similar redshift ($z\sim0.2$) and to the local Coma
sample of \cite{JFK95} (JFK95). In total, the low $L_{X}$ sample comprises
19 E+S0 galaxies, divided into 13 E and 6 S0 galaxies. On average, all these
objects show an evolution of $\overline{m}_{r}\sim0.24$ compared to Coma. The
34 rich cluster galaxies are on average brighter by $\sim$0.47 mag. The small,
not significant, difference might be caused by the fact, that the two samples
comprise slightly different galaxies in luminosity. However, as the scatter of
the low $L_{X}$ galaxies is quite large, a more quantitative analysis must be
based on the whole cluster sample.

\begin{acknowledgments}  
This project is a collaboration with R.~G.\ Bower, M.~L. Balogh and I. Smail
(Durham/UK) and R.~L.\ Davies (Oxford/UK) who have contributed to these results.
We thank the Calar Alto staff for efficient observational support.
AF and BLZ acknowledge financial support by the Volkswagen Foundation
(I/76\,520) and the DFG (ZI\,663/3-1, ZI\,663/5-1).
\end{acknowledgments}


\begin{thebibliography}{}


  \bibitem[Balogh et al. 2002a]{lowLxhst}
     {Balogh, M.~L., Smail, I., Bower, R.~G.} 2002a
     \textit{ApJ} \textbf{566}, 123--136. 

  \bibitem[Balogh et al. 2002b]{lowLx}
     {Balogh, M., Bower, R.~G., Smail, I., Ziegler, B.~L., Davies, R.~L.,
      Gaztelu, A., Fritz, A.} 2002b
     \textit{MNRAS} \textbf{337}, 256--274.
     
  \bibitem[Cole et al. 2000]{Col:00}
     {Cole, S., Lacey, C.~G., Baugh, C.~M., Frenk, C.~S.} 2000
     \textit{MNRAS} \textbf{319}, 168--204.
     
  \bibitem[Djorgovski \& Davis 1987]{DD87}
     {Djorgovski, S., Davis, M.} 1987
     \textit{ApJ} \textbf{313}, 59--68.

  \bibitem[Fritz et al. 2004]{Fri:04} 
     {Fritz, A., Ziegler, B.~L., Bower, R.~G. et al.} 2004
     \textit{MNRAS}, submitted.

  \bibitem[J{\o}rgensen et al. 1995]{JFK95}
     {J{\o}rgensen, I., Franx, M., Kj{\ae}rgaard, P.} 1995
     \textit{MNRAS} \textbf{273}, 1097--1128. (JFK95)

  \bibitem[van Dokkum \& Stanford 2003]{vDS03}
     {van Dokkum, P.~G., Stanford, S.~A.} 2003
     \textit{ApJ} \textbf{585}, 78--89.

  \bibitem[Ziegler \& Bender 1997]{ZB97}
     {Ziegler, B.~L., Bender, R.} 1997
     \textit{MNRAS} \textbf{291}, 527--543.

  \bibitem[Ziegler et al. 2001]{Z2001}
     {Ziegler, B.~L., Bower, R.~G., Smail, I. et al.} 2001
     \textit{MNRAS} \textbf{325}, 1571--1590.


\end{thebibliography}
\end{document}